\documentclass[final,5p,times,twocolumn]{elsarticle}

\usepackage{amssymb}

\usepackage{lineno}

\usepackage{color}
\usepackage{amsmath}

\journal{Physics Letters B}

\begin{document}

\begin{frontmatter}



\title{Search for two-neutrino double electron capture on $^{124}$Xe with the XMASS-I detector}

\address{\rm\normalsize XMASS Collaboration$^*$}
\cortext[cor1]{{\it E-mail address:} xmass.publications1@km.icrr.u-tokyo.ac.jp .}

\author[ICRR,IPMU]{K.~Abe}
\author[ICRR,IPMU]{K.~Hiraide}
\author[ICRR,IPMU]{K.~Ichimura}
\author[ICRR,IPMU]{Y.~Kishimoto}
\author[ICRR,IPMU]{K.~Kobayashi}
\author[ICRR]{M.~Kobayashi}
\author[ICRR,IPMU]{S.~Moriyama}
\author[ICRR]{K.~Nakagawa}
\author[ICRR,IPMU]{M.~Nakahata}
\author[ICRR]{T.~Norita}
\author[ICRR,IPMU]{H.~Ogawa}
\author[ICRR,IPMU]{H.~Sekiya}
\author[ICRR]{O.~Takachio}
\author[ICRR,IPMU]{A.~Takeda}
\author[ICRR,IPMU]{M.~Yamashita}
\author[ICRR,IPMU]{B.~S.~Yang}

\author[IBS]{N.~Y.~Kim}
\author[IBS]{Y.~D.~Kim}

\author[Gifu,TASAKA]{S.~Tasaka}

\author[IPMU,LIU]{J.~Liu}
\author[IPMU]{K.~Martens}
\author[IPMU]{Y.~Suzuki}

\author[Kobe]{R.~Fujita}
\author[Kobe]{K.~Hosokawa}
\author[Kobe]{K.~Miuchi}
\author[Kobe]{N.~Oka}
\author[Kobe]{Y.~Onishi}
\author[Kobe,IPMU]{Y.~Takeuchi}

\author[KRISS,IBS]{Y.~H.~Kim}
\author[KRISS]{J.~S.~Lee}
\author[KRISS]{K.~B.~Lee}
\author[KRISS]{M.~K.~Lee}

\author[Miyagi]{Y.~Fukuda}

\author[STE,KMI]{Y.~Itow}
\author[STE]{R.~Kegasa}
\author[STE]{K.~Kobayashi}
\author[STE]{K.~Masuda}
\author[STE]{H.~Takiya}
\author[STE]{H.~Uchida}

\author[Tokai1]{K.~Nishijima}

\author[YNU1]{K.~Fujii}
\author[YNU1]{I.~Murayama}
\author[YNU1]{S.~Nakamura}

\address[ICRR]{Kamioka Observatory, Institute for Cosmic Ray Research, the University of Tokyo, Higashi-Mozumi, Kamioka, Hida, Gifu, 506-1205, Japan}
\address[IBS]{Center of Underground Physics, Institute for Basic Science, 70 Yuseong-daero 1689-gil, Yuseong-gu, Daejeon, 305-811, South Korea}
\address[Gifu]{Information and multimedia center, Gifu University, Gifu 501-1193, Japan}
\address[IPMU]{Kavli Institute for the Physics and Mathematics of the Universe (WPI), the University of Tokyo, Kashiwa, Chiba, 277-8582, Japan}
\address[KMI]{Kobayashi-Maskawa Institute for the Origin of Particles and the Universe, Nagoya University, Furo-cho, Chikusa-ku, Nagoya, Aichi, 464-8602, Japan}
\address[Kobe]{Department of Physics, Kobe University, Kobe, Hyogo 657-8501, Japan}
\address[KRISS]{Korea Research Institute of Standards and Science, Daejeon 305-340, South Korea}
\address[Miyagi]{Department of Physics, Miyagi University of Education, Sendai, Miyagi 980-0845, Japan}
\address[STE]{Solar Terrestrial Environment Laboratory, Nagoya University, Nagoya, Aichi 464-8602, Japan}
\address[Tokai1]{Department of Physics, Tokai University, Hiratsuka, Kanagawa 259-1292, Japan}
\address[YNU1]{Department of Physics, Faculty of Engineering, Yokohama National University, Yokohama, Kanagawa 240-8501, Japan}

\fntext[TASAKA]{Now at Kamioka Observatory, Institute for Cosmic Ray Research, the University of Tokyo, Higashi-Mozumi, Kamioka, Hida, Gifu, 506-1205, Japan.}
\fntext[LIU]{Now at Department of Physics, the University of South Dakota, Vermillion, SD 57069, USA.}

\begin{abstract}
Double electron capture is a rare nuclear decay process in which
two orbital electrons are captured simultaneously in the same nucleus.
Measurement of its two-neutrino mode would provide a new reference
for the calculation of nuclear matrix elements whereas observation of its neutrinoless mode
would demonstrate lepton number violation.
A search for two-neutrino double electron capture on $^{124}$Xe is performed 
using 165.9 days of data collected with the XMASS-I liquid xenon detector.
No significant excess above background was observed and we set a lower limit
on the half-life as $4.7 \times 10^{21}$ years at 90\% confidence level.
The obtained limit has ruled out parts of some theoretical expectations.
We obtain a lower limit on the $^{126}$Xe two-neutrino double electron capture
half-life of $4.3 \times 10^{21}$ years at 90\% confidence level as well.
\end{abstract}


\end{frontmatter}



\section{Introduction}
The observed baryon asymmetry in the Universe still proves to be a fundamental challenge,
calling for physics beyond the standard model of particle physics.
Lepton number violation involving Majorana neutrinos
is one way to address this challenge in the context of leptogenesis~\cite{Buchmuller:2005eh}.
The most sensitive probe for lepton number violation is
neutrinoless double beta decay ($0\nu \beta^{-} \beta^{-}$)
\begin{equation}
    (Z,A) \to (Z+2,A) +2e^{-} \ ,
\end{equation}
where $Z$ and $A$ are the atomic number and atomic mass number of a given nucleus, respectively.
Its inverse, neutrinoless double electron capture ($0\nu$ECEC), is also a lepton number violating process
\begin{equation}
    (Z,A) +2e^{-} \to (Z-2,A) \ ,
\end{equation}
where two orbital electrons are captured simultaneously.
This process is expected to have a longer life-time and accompanied by a photon
that carries away the decay energy.
However, a possible enhancement of the capture rate by a factor as large as $10^{6}$
can occur if the masses of the initial and final (excited) nucleus are degenerate~\cite{Bernabeu:1983yb},
and hence this nuclear decay process is also attracting attention
both theoretically~\cite{Sujkowski:2003mb,Frekers:2005ze,Krivoruchenko:2010ng,Kotila:2014zya}
and experimentally~\cite{Barabash:2006qx,Barabash:2009ja,Belli:2013qja,Belli:2014map,Gavrilyuk:2013yqa}.
Moreover, neutrinoless positron-emitting electron capture ($0\nu \beta^{+}$EC)
and neutrinoless double beta plus decay ($0\nu \beta^{+} \beta^{+}$) may occur in the same nucleus
depending on the mass difference between the initial and final nuclei.
Detection of these nuclear decay modes could help to determine the effective neutrino mass
and parameters of a possible right-handed weak current~\cite{Haxton:1985am,Hirsch:1994es}.

On the other hand, two-neutrino double beta decay (2$\nu \beta^{-} \beta^{-}$) and
two-neutrino double electron capture (2$\nu$ECEC) processes are allowed within
the standard model.
Although 2$\nu \beta^{-} \beta^{-}$ has been observed in more than ten isotopes,
there exist only a few positive experimental results for 2$\nu$ECEC so far:
a geochemical measurement for $^{130}$Ba with a half-life of
$(2.2\pm 0.5)\times 10^{21}$ years~\cite{Meshik:2001ra}
and a direct measurement for $^{78}$Kr with a half-life of 
$(9.2^{+5.5}_{-2.6}(stat)\pm 1.3 (sys))\times 10^{21}$ years~\cite{Gavrilyuk:2013yqa}.
In the case that after the 2$\nu$ECEC process the nucleus is in the ground state,
the observable energy comes from atomic de-excitation and nuclear recoil;
depending on the nucleus, the energy deposited by nuclear recoil may become 
negligible, leading to a well defined energy deposit dominated by the 
atomic de-excitation - a line spectrum.
Nevertheless, little attention has been paid to direct detection
of this process because of difficulties due to small natural abundance and
the energy threshold of large volume detectors.
Any measurement of 2$\nu$ECEC will provide a new reference for the calculation
of nuclear matrix elements
from the proton-rich side of the mass parabola of
even-even isobars~\cite{Suhonen:1998ck}.
Although the matrix element for the two-neutrino mode is different from
that for the neutrinoless mode, it gives constraints on the relevant parameters
within a chosen model~\cite{Bilenky:2014uka}. 

The XMASS detector uses liquid xenon in its natural isotopic abundance
as its active target material. Among others it contains
the double electron capture nuclei $^{124}$Xe (0.095\%) and $^{126}$Xe (0.089\%),
as well as the double beta decay nuclei $^{136}$Xe (8.9\%) and $^{134}$Xe (10.4\%).
It has been pointed out that large volume dark matter detectors
with natural xenon as targets have the potential to measure the 2$\nu$ECEC
on $^{124}$Xe~\cite{Mei:2013cla,Barros:2014exa}. 
Among the different models for calculating the corresponding nuclear matrix element,
there exists a wide spread of calculated half-lives for this process:
between $10^{20}$ and $10^{24}$~years as summarized in Table~\ref{table:calculated_halflives}.

\begin{table}[t]
 \caption{Calculated half-lives for 2$\nu$ECEC on $^{124}$Xe.
 The lower and upper values are calculated for the axial-vector coupling constant $g_A=1.26$ and 1.0, respectively.}
 \label{table:calculated_halflives}
 \begin{center}
  \begin{tabular}{lll}
   \hline \hline
   Model & $T_{1/2}^{2\nu{\rm ECEC}}$ ($\times 10^{21}$ yr) & Reference \\ \hline
   QRPA                  & 0.4-8.8 & ~\cite{Suhonen:2013rca} \\
   QRPA                  & 2.9-7.3 & ~\cite{Hirsch:1994es} \\
   SU(4)$_{\sigma \tau}$ & 7.0-18  & ~\cite{Rumyantsev:1998uy} \\
   PHFB                  & 7.1-18  & ~\cite{Singh:2007jh} \\
   PHFB                  & 61-160  & ~\cite{Shukla:2007ju} \\
   MCM                   & 390-980 & ~\cite{Aunola:1996ui} \\
   \hline \hline
  \end{tabular}
 \end{center}
\end{table}

A previous experiment used enriched xenon.
A gas proportional counter containing 58.6~g of $^{124}$Xe (enriched to 23\%) was
looking for the simultaneous capture of two $K$-shell electrons on that isotope,
and published the latest lower bound on the half-life T$_{1/2}^{2\nu2K}$ as
$2.0\times 10^{21}$~years~\cite{Gavrilyuk:2014dqa,Gavrilyuk:2015ada}.

In this paper, we present the result from a search for 2$\nu$ECEC on $^{124}$Xe
using the XMASS-I liquid xenon detector.

\section{The XMASS-I Detector}
XMASS-I is a large single phase liquid xenon detector~\cite{Abe:2013tc} located
underground (2700\,m water equivalent) at the Kamioka Observatory in Japan. 
An active target of 835\,kg of liquid xenon is held inside of a
pentakis-dodecahedral copper structure that holds 642 inward-looking 
photomultiplier tubes (PMTs) on its approximately spherical inner surface. 
The detector is calibrated regularly with $^{57}$Co and $^{241}$Am sources~\cite{Kim:2015rsa}
inserted along the central vertical axis of the detector. 
Measuring with the $^{57}$Co source from the center of the detector volume 
the photoelectron yield is determined to be
13.9~photoelectrons (PEs)/keV~\cite{FN1}.
This large photoelectron yield is realized
by a large inner surface photocathode coverage of $>$62\%
and the large PMT quantum efficiency of approximately 30\%.
The non-linear response in scintillation light yield for
electron-mediated events in the detector was calibrated with $^{55}$Fe,
$^{57}$Co, $^{109}$Cd, and $^{241}$Am sources.
When a PMT signal exceeds the discriminator threshold equivalent to 0.2~PE,
a ``hit'' is registered on the channel.
Data acquisition is triggered if ten or more hits are asserted within 200\,ns.
Each PMT signal is digitized with charge and timing resolution of
0.05\,PE\ and 0.4\,ns, respectively~\cite{Fukuda:2002uc}.
The liquid xenon detector is located at the center of a cylindrical water Cherenkov
veto counter and shield, which is 11\,m high with a 10\,m diameter.
The veto counter is equipped with 72 20-inch PMTs.
Data acquisition for the veto counter is triggered if eight or more of its 
PMTs register a signal within 200\,ns.
XMASS-I is the first direct detection dark matter experiment
equipped with such an active water Cherenkov shield.

\section{Expected Signal and Detector Simulation}
The process of 2$\nu$ECEC on $^{124}$Xe is
\begin{equation}
    ^{124}{\rm Xe} + 2e^- \to ^{124}{\rm Te} + 2\nu_e
\end{equation}
with a $Q$-value of 2864 keV.
In the case that two $K$-shell electrons in the $^{124}$Xe atom are captured simultaneously,
a daughter atom of $^{124}$Te is formed with two vacancies in the $K$-shell and
de-excites by emitting atomic X-rays and/or Auger electrons.
The total energy deposition in the detector is $2K_{b} = 63.63$~keV,
where $K_{b}$ is the binding energy of a $K$-shell electron in a tellurium atom.
The energy deposition from the recoil of the daughter nucleus is $\sim$30~eV
at most, which is negligible.
Although $^{126}$Xe can also undergo 2$\nu$ECEC,
this reaction is expected to be much slower than that on $^{124}$Xe
since its $Q$-value of 920~keV is smaller.
The $Q$-values are taken from the AME2012 atomic mass evaluation~\cite{AME2012}.

The Monte Carlo (MC) generation of the atomic de-excitation signal is based on
the atomic relaxation package in Geant4 \cite{2007ITNS...54..585G}.
While the X-ray and Auger electron tables refer to emission from singly charged ions,
2$\nu$ECEC produces a doubly charged ion.
The energy of the double-electron holes in the $K$-shell of $^{124}$Te is calculated
to be 64.46~keV~\cite{Nesterenko:2012xp}, which is only 0.8~keV different from the sum of the
$K$-shell binding energy of the singly charged ion.
Therefore, this difference is negligible in this analysis.
Simulated de-excitation events are generated uniformly throughout the detector volume. 
The MC simulation includes the nonlinearity of the scintillation response \cite{Abe:2013tc}
as well as corrections derived from detector calibrations.
The absolute energy scale of the MC is adjusted at 122\,keV.
The systematic difference of the energy scale
between data and MC due to imperfect modeling of the nonlinearity in MC
is estimated as 3.5\% by comparing $^{241}$Am data to MC.
The decay constants of scintillation light and the timing response of the PMTs are
modeled to reproduce the time distribution observed with the $^{57}$Co (122\,keV) and
$^{241}$Am (60\,keV) gamma ray sources \cite{Uchida:2014cnn}.
The group velocity of the scintillation light in liquid xenon is
calculated from its refractive index ($\sim$11\,cm/ns for 175\,nm) 
\cite{LXeSpeed}.

\section{Data Sample and Event Selection}
The data used in the present analysis were collected between December 24, 2010 and May 10, 2012.
Since we took extensive calibration data and various special runs
by changing the detector conditions to understand the general detector response and the background, we
select periods of operation under what we call normal data taking conditions
with a stable temperature ($174\pm 1.2$~K) and pressure (0.160-0.164~MPa absolute).
After furthermore removing periods of operation with excessive PMT noise, unstable pedestal levels,
or abnormal trigger rates, the total livetime becomes 165.9~days.

Event selection proceeds in four stages: pre-selection, fiducial volume cut,
timing balance cut, and band-like pattern cut.
The pre-selection requires that no outer detector trigger is associated with the event,
that the event is separated in time from the nearest event by at least 10~ms,
and that the RMS spread of the inner detector hit timings of the event is less than 100~ns.
This pre-selection reduces the total effective lifetime
to 132.0 days in the final sample.

%

In order to select events occurring in the fiducial volume,
an event vertex is reconstructed based on a maximum likelihood evaluation of
the observed light distribution in the detector~\cite{Abe:2013tc}.
We select events satisfying that the radial distance of their reconstructed
vertex from the center of the detector is smaller than 15~cm.
The fiducial mass of natural xenon in that volume is 41~kg,
containing 39~g of $^{124}$Xe.

During the data-taking period, a major background in the relevant energy range comes from
radioactive contaminants in the aluminum seal of the PMTs. These background events
often occur at a blind corner of the nearest PMT and are mis-reconstructed in the
inner volume of the detector. 
The remaining two cuts deal with these mis-reconstructed events.
The timing balance cut uses the time difference between the first hit
in an event and the mean of the timings of the second half of
all the time-ordered hits in the event.
Events with smaller time difference are less likely to be events from the detector's
inner surface that were wrongly reconstructed and are kept.
The timing balance cut reduces the data by a factor of 5.9 in the signal energy window
defined later, while it keeps 80\% of signal events remaining after the fiducial volume cut.
The band-like pattern cut eliminates events that reflect their origin within grooves or crevices
in the inner detector surface through a particular illumination pattern:
The rims of the groove or crevice act as an aperture that is
projected as a ``band'' of higher photon counts onto the inner detector surface.
This band is characterized by the ratio of the maximum PEs in the band of width 15~cm
to the total PEs in the event~\cite{Uchida:2014cnn}.
Events with smaller ratio are less likely to originate from crevices and are selected.
The band pattern cut reduces the data by a factor of 24.6 while it keeps 70\% of signal events
remaining after the fiducial volume and timing balance cuts.

\begin{figure}[tp]
  \begin{center}
    \includegraphics[keepaspectratio=true,height=60mm]{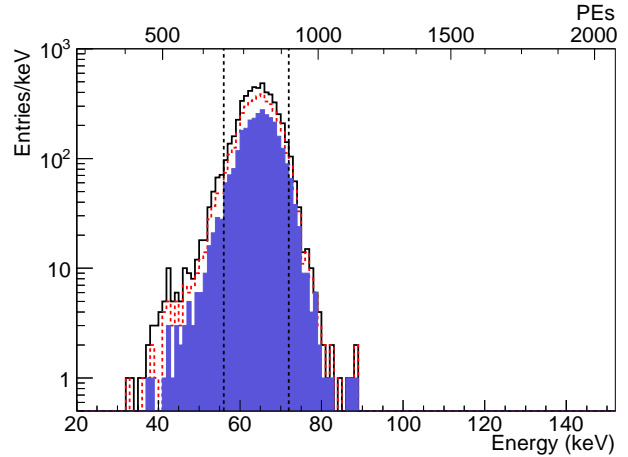}
  \end{center}
  \caption{Energy spectra of the simulated events after each reduction step.
  From top to bottom, the simulated energy spectrum after pre-selection and radius cut (black solid),
  timing balance cut (red dashed), and band-like pattern cut (blue filled) are shown.
  The vertical dashed lines indicate the 56-72~keV signal window.}
  \label{fig:mc-reduction}
\end{figure}

The fiducial volume, timing balance, and band pattern cut values are optimized
to maximize sensitivity to a monoenergetic peak in the 60 keV region.
For the fiducial volume cut, the range of the cut value was restricted in the
optimization process to be larger than 15~cm in order to avoid too small of
an acceptance, and this restriction turns out to determine the optimal value~\cite{Abe:2014zcd}.

In the present analysis, the total energy deposition of events is reconstructed
from the observed number of photoelectrons correcting for the non-linear response of scintillation light yield.
The correction is performed assuming the light originates from two X-rays with equal energy.
Finally, the signal window is defined such that it contains 90\% of the simulated
signal with equal 5\% tails to either side after all the above were applied,
which results in a 56$-$72~keV window. 
Fig.~\ref{fig:mc-reduction} shows energy spectra of the simulated events after each reduction step.
From the simulation, signal detection efficiency is estimated to be 59.7\%.

\section{Results and Discussion}

\begin{figure}[tbp]
  \begin{center}
    \includegraphics[keepaspectratio=true,height=60mm]{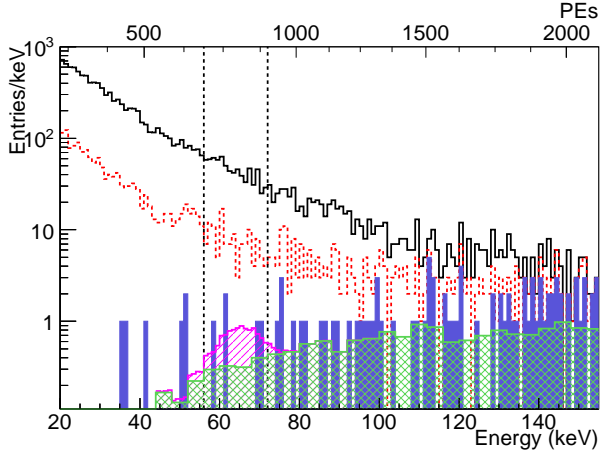}
  \end{center}
  \caption{Energy spectra of the observed events after each reduction step for the 165.9 days of data.
  From top to bottom, the observed energy spectrum after pre-selection and radius cut (black solid), timing balance cut (red dashed),
  and band-like pattern cut (blue filled) are shown. The vertical dashed lines indicate the 56-72~keV signal window.
  The expected $^{214}$Pb background (green hatched) together with the signal expectation for
  the 90\% confidence level upper limit (magenta hatched) are also shown.}
  \label{fig:data-reduction}
\end{figure}

Fig.~\ref{fig:data-reduction} shows energy distributions of data events remaining after each reduction step.
After all cuts, 5 events are left in the signal region but no significant peak is seen.
The main contribution to the remaining background in this energy regime is the $^{222}$Rn daughter $^{214}$Pb
in the detector. The amount of $^{222}$Rn was estimated to be $8.2\pm 0.5$~mBq from the observed rate of
$^{214}$Bi-$^{214}$Po consecutive decays.
Given the measured decay rate the expected number of background events in the signal region
from this decay alone is estimated to be $5.3\pm 0.5$ events.
The concentration of krypton in the xenon was measured to be $<$2.7~ppt~\cite{Abe:2013tc},
and thus background from $^{85}$Kr is negligible in this analysis.
The background from $2\nu \beta^{-} \beta^{-}$ of $^{136}$Xe ($T_{1/2}=2\times 10^{21}$ years~\cite{Agashe:2014kda})
is smaller than the $^{214}$Pb background by a factor of 7 and is negligible for this analysis.

\begin{figure}[tbp]
  \begin{center}
    \includegraphics[keepaspectratio=true,height=60mm]{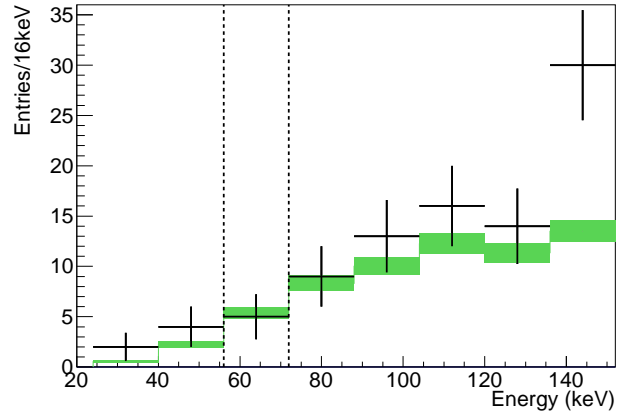}
  \end{center}
  \caption{Energy distribution of the observed events (black) overlaid with the $^{214}$Pb background
  simulation (green) after all cuts except for the energy window cut.
  The vertical dashed lines indicate the 56-72~keV signal window.}
  \label{fig:data-bg-overlay}
\end{figure}

Fig.~\ref{fig:data-bg-overlay} shows the energy distribution of the observed events overlaid
with the $^{214}$Pb background simulation after all cuts except for the energy window cut.
The energy spectrum after cuts in data is consistent with the expected $^{214}$Pb background spectrum.
Although the total number of observed events is 26\% larger than that of the expected $^{214}$Pb background
in the energy range between 24~keV and 136~keV but outside the signal window,
the tension is still at a 1.4$\sigma$ level with this small statistics.
Note that an excess in the highest energy bin is due to a gamma-ray from $^{\rm 131m}$Xe in liquid xenon,
and thus this energy bin is not included in the calculation. 
We derive a conservative limit under the assumption of the $^{214}$Pb background constrained
by the $^{214}$Bi-$^{214}$Po measurement.

A lower limit on the 2$\nu$ECEC half-life
is derived using the following Bayesian method that also accounts for systematic uncertainties
to calculate the conditional probability distribution for the decay rate as follows:
\begin{eqnarray}
   P (\Gamma| n_{\rm obs}) &=& \iiiint \frac{{\rm e}^{-(\Gamma \lambda \epsilon +b) (1+\delta)}
   ((\Gamma \lambda \epsilon +b) (1+\delta))^{n_{\rm obs}}}{n_{\rm obs}!} \nonumber \\
   & & \times  P(\Gamma) P(\lambda) P(\epsilon) P(b) P(\delta)
   {\rm d} \lambda {\rm d} \epsilon {\rm d} b {\rm d} \delta
\end{eqnarray}
where $\Gamma$ is the decay rate, $n_{\rm obs}$ is the observed number of events,
$\lambda$ is the detector exposure including the abundance of $^{124}$Xe,
$\epsilon$ is the detection efficiency, $b$ is the expected number of background events,
and $\delta$ is a parameter representing the systematic uncertainty
in the event selection which affects both signal and background.
The decay rate prior probability $P(\Gamma)$ is 1 for $\Gamma \ge 0$ and otherwise 0. 
The prior probability distributions incorporating systematic uncertainties
in the detector exposure $P(\lambda)$, detection efficiency $P(\epsilon)$, background $P(b)$,
and event selection $P(\delta)$ are assumed to be the split normal distribution centered
at the nominal value with two standard deviations since some error sources are
found to have a different impact on the positive versus the negative side of the distribution center
as described below.

Table~\ref{table:systematics} summarizes the systematic uncertainties in exposure,
detection efficiency, and event selection.
The systematic uncertainty in the detector exposure is dominated by the uncertainty
in the abundance of $^{124}$Xe in the xenon. A sample
was taken from the detector and its isotope composition was measured
at the Geochemical Research Center, the University of Tokyo
using a modified VG5400/MS-III mass spectrometer~\cite{Bajo:2011}.
The result is consistent with that of natural xenon in air, and
we treat the uncertainty in that measurement as a systematic error.
The systematic uncertainty in the detection efficiency is estimated from comparisons
between data and MC simulation for $^{241}$Am (60~keV $\gamma$-ray) calibration data
at various positions within the fiducial volume.
The systematic uncertainty in the energy scale is evaluated to be $\pm 5\%$,
summing up in quadrature the uncertainties from the nonlinearity of the scintillation
yield ($\pm 3.5\%$), position dependence ($\pm 2\%$), and time variation ($\pm 3\%$).
Changing the number of photons generated per unit energy deposited in the simulation
by this amount, the signal efficiency changes by $\pm ^{0}_{8.6}\%$.
Since we apply the energy cut on lower and upper sides,
both increasing and decreasing number of photons in MC makes signal efficiency smaller.
The energy resolution in the calibration data is found to be
12\% worse than that in the simulation. The uncertainty due to this
difference is evaluated by worsening energy resolution in the simulation,
which leads to a $5.3\%$ reduction in signal efficiency.

The uncertainty in modeling the scintillation decay constant as a function of
energy is evaluated to be $\pm ^{1.5}_{0}$~ns, resulting in an uncertainty in the
signal efficiency of $\pm ^{0}_{7.1}\%$.
The radial position of the reconstructed vertex for the calibration data differs
from the true source position by 5~mm, which causes a $6.7\%$ reduction in efficiency.
For the timing balance and band-like pattern cuts, we evaluate the impact on the signal efficiency
by again taking the difference of their acceptance for calibration data and the respective simulation.
The resulting change in signal efficiency is $\pm ^{3.0}_{0}\%$ for the timing balance cut,
and $\pm 5.0\%$ for the band-like pattern cut.

\begin{table}[tbp]
 \caption{Summary of systematic uncertainties in exposure, detection efficiency, and event selection.}
 \label{table:systematics}
 \begin{center}
  \begin{tabular}{llc}
    \hline \hline
    Item & Error source & Fractional \\
         &              & uncertainty (\%) \\
    \hline
    Exposure   & Abundance of $^{124}$Xe  &  $\pm 8.5$ \\
               & Liquid xenon density     &  $\pm 0.5$ \\
    \hline
    Efficiency & Energy scale             &  $\pm ^{0}_{8.6}$ \\
               & Energy resolution        &  $\pm ^{0}_{5.3}$ \\
               & Scintillation decay time & $\pm ^{0}_{7.1}$ \\
    \hline
    Event selection & Fiducial volume cut   & $\pm ^{0}_{6.7}$ \\
                    & Timing balance cut    & $\pm ^{3.0}_{0}$ \\
                    & Band-like pattern cut & $\pm 5.0$ \\
    \hline \hline
  \end{tabular}
 \end{center}
\end{table}

Finally, we calculate the 90\% confidence level (CL) limit using the relation
\begin{equation}
\frac{\int_{0}^{\Gamma_{\rm limit}} P(\Gamma|n_{\rm obs}) {\rm d} \Gamma}
{\int_{0}^{\infty} P(\Gamma|n_{\rm obs}) {\rm d} \Gamma} = 0.9
\end{equation}
to obtain
\begin{equation}
T_{1/2}^{2\nu 2K} \left (^{124}{\rm Xe} \right ) = \frac{\ln 2}{\Gamma_{\rm limit}} > 4.7 \times 10^{21} \ {\rm years}.
\end{equation}
Note that the total systematic uncertainty worsens the obtained limit by 20\%.

In addition, the fact that we do not observe significant excess above background allows us to give
a constraint on 2$\nu$ECEC on $^{126}$Xe in the same manner.
The fiducial volume contains 36~g of $^{126}$Xe and the uncertainty in the abundance of $^{126}$Xe is
estimated to be 12.1\%, and we obtain
$T_{1/2}^{2\nu 2K} \left (^{126}{\rm Xe} \right ) > 4.3 \times 10^{21} \ {\rm years}$
at 90\% CL.

The XMASS project uses a single phase liquid xenon detector with a natural abundance target.
This straightforward technology offers easy scalability to larger detectors. 
The future XMASS-II detector will contain 10~tons of liquid xenon in its fiducial volume
as target, and the expected sensitivity of XMASS-II will improve by more than two orders
of magnitude over the current limit after 5 years,
assuming a background level of $3\times 10^{-5}$ day$^{-1}$kg$^{-1}$keV$^{-1}$.
This background is due to 2$\nu \beta^{-} \beta^{-}$ of $^{136}$Xe and $pp$+$^{7}$Be solar neutrinos.

\section{Conclusions}
In conclusion, we have searched for 2$\nu$ECEC on $^{124}$Xe
using an effective live time of 132.0 days of XMASS-I data 
in a fiducial volume containing 39~g of $^{124}$Xe.
No significant excess over the expected background is found in the signal region,
and we set a lower limit on its half-life of $4.7 \times 10^{21}$ years at 90\% CL.
The obtained limit has ruled out parts of some theoretical expectations.
In addition, we obtain a lower limit on the $^{126}$Xe 2$\nu$ECEC half-life of
$4.3 \times 10^{21}$ years at 90\% CL.
A future detector with XMASS-II characteristics establishes a path toward
covering the whole range of half-lives obtained
in the model calculations cited in introduction.

\section*{Acknowledgments}
We thank Prof. Nagao and his group for the measurement of the isotope
composition of the xenon used in the XMASS-I detector.
We gratefully acknowledge the cooperation of the Kamioka Mining and Smelting Company. 
This work was supported by the Japanese Ministry of Education,
Culture, Sports, Science and Technology, Grant-in-Aid for Scientific Research, 
JSPS KAKENHI Grant No. 19GS0204 and 26104004,
and partially by the National Research Foundation of Korea Grant funded
by the Korean Government (NRF-2011-220-C00006).





\end{document}